\documentclass[preprint,12pt]{elsarticle}

\usepackage{verbatim} 
\usepackage{epsfig}
\usepackage{amssymb} 
\usepackage{mathtools}
\usepackage{lineno}
\usepackage{tikz}        
\usepackage{etoolbox}
\usepackage{ulem}         
\usepackage[caption=false]{subfig}
\usepackage{textcomp}

\begin{document}

\newcommand*{\hwplotB}{\raisebox{3pt}{\tikz{\draw[red,dashed,line 
width=3.2pt](0,0) -- 
(5mm,0);}}}

\newrobustcmd*{\mydiamond}[1]{\tikz{\filldraw[black,fill=#1] (0,0) -- 
(0.1cm,0.2cm) --
(0.2cm,0) -- (0.1cm,-0.2cm);}}

\newrobustcmd*{\mytriangleleft}[1]{\tikz{\filldraw[black,fill=#1] (0,0.15cm) -- 
(-0.3cm,0) -- (0,-0.15cm);}}
\definecolor{Blue}{cmyk}{1.,1.,0,0} 

\begin{frontmatter}

\title{Social Force Model parameter testing and optimization using a high 
stress real-life situation}

\author[add1]{I.M.~Sticco}
 \address[add1]{Departamento de F\'\i sica, Facultad de Ciencias 
Exactas y Naturales, \\ Universidad de Buenos Aires,\\
 Pabell\'on I, Ciudad Universitaria, 1428 Buenos Aires, Argentina.}

 \author[add2]{G.A.~Frank}
 \address[add2]{Unidad de Investigaci\'on y Desarrollo de las 
Ingenier\'\i as, Universidad Tecnol\'ogica Nacional, Facultad Regional Buenos 
Aires, Av. Medrano 951, 1179 Buenos Aires, Argentina.}

\author[add1,add3]{C.O.~Dorso\corref{cor1}}%
 \cortext[cor1]{codorso@df.uba.ar}

 \address[add3]{Instituto de F\'\i sica de Buenos Aires,\\
Pabell\'on I, Ciudad Universitaria, 1428 Buenos Aires, Argentina.}

\begin{abstract}

The escape panic version of the Social Force Model (SFM) is a suitable model 
for describing emergency evacuations. In this research, we analyze a real-life 
video, recorded at the opening of a store during a Black Friday event, which 
resembles an emergency evacuation (November 2017, 
South Africa). We measure 
the flow of pedestrians entering the store and found a higher value ($\langle J  
\rangle =6.7 \pm 0.8\,$p/s) than the usually reported in ``laboratory'' 
conditions. We performed numerical simulations to recreate this 
event. The empirical measurements were compared against simulated evacuation 
curves corresponding to different sets of parameters currently in use in the 
literature. The results 
obtained suggest that the set of parameters corresponding to calibrations 
from laboratory experiments (involving pedestrians in which the safety of the 
participants is of major concern) or situations where the physical contact is 
negligible, produce simulations in which the agents evacuate faster than in 
the empirical scenario. 
To conclude the paper, we optimize two parameters of the model: the friction 
coefficient $k_t$ and the body force coefficient $k_n$. The best fit we found 
could replicate the qualitative and quantitative behavior of the empirical 
evacuation curve. We also found that many different combinations in the 
parameter space can produce similar results in terms of the goodness of fit.\\

\end{abstract}


\end{frontmatter}


\section{\label{introduction}Introduction}

In the last years, a growing interest appeared in the research of 
pedestrian dynamics and emergency evacuations. The increasing number of 
tragedies in night clubs~\cite{veltfort1943cocoanut,atiyeh2013brazilian}, 
sports events~\cite{wright1993recall,delaney2002mass}, and mass gatherings in 
general~\cite{illiyas2013human,ngai2009human}, has concerned many people around 
the world. The urgency for understanding and preventing these tragedies has 
called the attention of researchers from different disciplines.\\

There are three main approaches to study evacuations and pedestrian dynamics in 
general: the theoretical 
approach~\cite{helbing1995social,henderson1974fluid,blue1999cellular}, 
the experimental 
approach ($i.e$ performing 
laboratory experiments)~\cite{pastor2015experimental,liao2014experimental,
nicolas2017pedestrian}, and the empirical 
approach ($i.e$ analyzing 
real-life data)~\cite{corbetta2014high,zhang2013empirical}. \\

The emergency evacuation in a panic state is a very complex phenomenon that 
requires the three approaches to be properly addressed. Each of these approaches 
has its strengths and limitations. The laboratory experiments (experimental 
approach) allow the researcher to have control over the analyzed situation while 
recreating some of the traits that occur in a real-life scenario. However, 
doing experiments that truly reproduce emergency 
evacuations is impossible due to safety reasons. The analysis of real-life 
situations (empirical approach) gives the researcher the 
possibility of studying the system with a closer degree of realism. However, 
the limitation of this approach is that there are not many documented reports 
on real-life emergency evacuations yet.\\

The theoretical approach is the most common one. This approach does not have to 
deal with the difficulties of the experimental and empirical approaches. 
Nevertheless, the theoretical models may produce unrealistic results if they 
are not validated or adequately calibrated. \\

Many pedestrian dynamics models were proposed in the last decades (see 
Refs.~\cite{chen2018social,schadschneider2009fundamentals}). There 
are two main groups of pedestrian models: the discrete 
models~\cite{blue1999cellular,fukui1999self} in which 
space and time are discretized and the continuous 
models~\cite{helbing_2000} which consider space and time as continuous 
variables. The Social Force 
Model~\cite{helbing1995social,helbing_2000} belongs to the 
group of continuous models and it is one 
of the most developed models in pedestrian dynamics. Since its first version, 
there have been many revisions, extensions, and improvements to 
solve multiple problems that go beyond the emergency 
evacuations~\cite{koster2013avoiding,chen2018social}.\\

In this paper, we analyze a real-life video (empirical approach) of a Black 
Friday event where pedestrians rush to enter into a Nike store. The video 
was specially selected because the crowd behavior resembles an emergency 
evacuation. 
We also perform numerical 
simulations (theoretical approach) to recreate the results of the empirical 
scenario. We are only focusing on the escape panic version of the social 
force model~\cite{helbing_2000} because this version is specially 
oriented to reproduce emergency evacuations.\\

The paper is organized as follows. In the following section, we describe the 
forces and the proposed set of parameters that involve the escape panic version 
of the social force model. In section~\ref{model}, we mention the research 
scope and clarify some concepts that will be useful throughout the paper. In 
section~\ref{scope}, we provide details of the empirical measurements and the 
numerical simulations that we performed. Section~\ref{results} shows the 
results from both the empirical and numerical analyses. In the last part, we 
summarize the main conclusions of the work. 

\section{\label{model} The escape panic version of the Social Force Model}

\subsection{The forces of the model}

The escape panic version of the social force model came up for the first time in 
the year 2000~\cite{helbing_2000}. Unlike other 
versions~\cite{helbing1995social,johansson2007specification}, this one reduces 
the force description of the pedestrian evacuations to just a small number of 
parameters.\\

The dynamic of the simulation is determined by an equation of motion 
involving both “socio-psychological” forces and physical forces. The equation 
of motion for any pedestrian $i$ (of mass $m_i$) reads

\begin{equation}
 m_i\,\displaystyle\frac{d\mathbf{v}_i}{dt}=\mathbf{f}_d^{(i)}+
 \displaystyle\sum_{j=1}^N\mathbf{f}_s^{(ij)}+
 \displaystyle\sum_{j=1}^N\mathbf{f}_p^{(ij)}\label{newton_ec}
\end{equation}

\noindent where the subscript $j$ corresponds to the neighboring pedestrians or
the surrounding walls. The desired force $\mathbf{f}_d$ represents the
acceleration (or deceleration) of the pedestrian due to his (her) own will.  The 
social force $\mathbf{f}_s$, instead, describes the tendency of the pedestrians 
to stay away from each other. The physical force $\mathbf{f}_p$ stands for both 
the sliding friction and the body force repulsion. These forces are essential 
for understanding the particular effects of panicking crowds. \\

The pedestrians' own will is modeled by the desired force $\mathbf{f}_d$.  This 
force stands for the acceleration (deceleration) required to move at the desired 
velocity $v_d$. This parameter is associated with the anxiety level to reach a 
specific target. The parameter $\tau$ is a characteristic time scale that 
reflects the reaction time. Thus, the desired force is modeled as follows

\begin{equation}
\mathbf{f}_d^{(i)}=m\,\displaystyle\frac{v_d^{(i)}\,
\hat{\mathbf{e}}_d^{(i)}(t)-
 \mathbf{v}^{(i)}(t)}{\tau}
\end{equation}

\noindent where $\hat{\mathbf{e}}(t)$ represents the unit vector pointing to the
target position and $\mathbf{v}(t)$ stands for the pedestrian velocity at time 
$t$.
\\

The tendency of any individual to preserve his (her) ``private sphere'' is
accomplished by the social force $\mathbf{f}_s$. Therefore, $\mathbf{f}_s$ 
reflects the psychological tendency of two pedestrians to stay away from each 
other.  The mathematical expression to model this kind of 
``socio-psychological'' behavior is as follows

\begin{equation}
 \mathbf{f}_s^{(i)}=A\,e^{(R_{ij}-r_{ij})/B}\,\hat{\mathbf{n}}_{ij}
 \label{eqn_social}
\end{equation}

\noindent where $r_{ij}$ means the distance between the center of mass of the
pedestrians $i$ and $j$, and $R_{ij}=R_i+R_j$ is the sum of the pedestrian's 
radius. The unit vector $\hat{\mathbf{n}}_{ij}$ points from pedestrian $j$ to 
pedestrian $i$, meaning a repulsive interaction.\\

The parameter $B$ is a characteristic scale that plays the role of a fall-off 
length within the social repulsion. Besides, the
parameter $A$ represents the intensity of the social repulsion. \\

The  expression for the physical force (say, the sliding 
friction plus the body force) has been borrowed from the granular matter 
field; the mathematical expression reads as follows

\begin{equation}
 \mathbf{f}_p^{(ij)}=k_t\,g(R_{ij}-r_{ij})\,
(\Delta\mathbf{v}^{(ij)}\cdot\hat{\mathbf{t}}_{ij})\,\hat{\mathbf{t}}_{ij}+
k_n\,g(R_{ij}-r_{ij})\,
\,\hat{\mathbf{n}}_{ij}\label{eqn_physical}
\end{equation}

\noindent where $g(R_{ij}-r_{ij})$ equals $R_{ij}-r_{ij}$ if $R_{ij}>r_{ij}$ and
vanishes otherwise. $\Delta\mathbf{v}^{(ij)}\cdot\hat{\mathbf{t}}_{ij}$
represents the relative tangential velocities of the sliding  bodies (or between
the individual and the walls). Further details can be found in 
Ref.~\cite{helbing_2000}. \\

The sliding friction occurs in the tangential direction $\hat{\mathbf{t}}_{ij}$ 
while the body force
occurs in the normal direction $\hat{\mathbf{n}}_{ij}$. Both forces are assumed 
to be linear with respect to the net distance between contacting pedestrians 
(overlap). 
The sliding friction is also linearly related to the difference between the 
(tangential) velocities. The parameter $k_t$ is a friction coefficient, while 
the parameter $k_n$ is a stiffness coefficient (analogous to the constant factor 
in the Hooke's law). For simplicity, we will omit the units of the parameters 
$k_n$ and $k_t$. Remember that the friction coefficient
has units $\left [ k_t \right]=$Kg$\,$m$^{-1}\,$s$^{-1}$ and
the body stiffness coefficient $\left [ k_n \right]=$Kg$\,$s$^{-2}$. \\

\subsection{The proposed sets of parameters for the model}

The escape panic version of the SFM involves 8 parameters: $m_i$, $R_i$, 
$\tau$, $v_d$, $A$, $B$, $k_n$ and $k_t$. The original version of the model 
fixes the mass $m_i=80\,$kg, $R_i$ was uniformly distributed in the interval 
[0.25\,m, 0.35\,m], to represent average soccer fans. The parameters 
$A=2000\,$N, $B=0.08\,$m take these values to reproduce the distance kept at 
normal desired velocities and the measured flows through 
bottlenecks~\cite{weidmann1993transporttechnik}. $\tau=0.5\,$s is said to be a 
reasonable estimate for the acceleration time. Regarding the parameters of the 
physical forces, $k_n= 1.2 \times 10^{5}$ and $k_t= 2.4 
\times 10^{5}$, Ref.~\cite{helbing_2000} does not express reasons 
for choosing these values, but $k_n$ may be related to the 
research in Ref.~\cite{melvin1988aatd}. Although this set of parameters has 
been used in many research 
studies~\cite{ha2012agent,song2018characteristic,saboia2012crowd,yang2014guided,
zhang2016simulation,zhao2017optimal,parisi2005microscopic}, there are other sets 
of parameters that have also been proposed to fit specific situations.\\

In Ref.~\cite{li_2015}, Li et al. calibrate the parameters $A$, $v_d$, $k_t$, 
$k_n$ from a real-life video analysis. The situation studied was an emergency 
evacuation of a classroom during the 2013 Ya\textquotesingle an  earthquake in 
China~\cite{video_yaan}. They use a Differential Evolution algorithm to fit the 
numerical simulation results to the empirical evacuation curve (cumulative 
number of evacuees vs. time). \\

In the research conducted by Haghani et al.~\cite{haghani2019simulating}, the 
authors perform a sensitivity analysis where they find that the parameters 
$\tau$ and $k_t$ are the most sensitive (the simulation's output is strongly 
dependent on the value of these parameters). They also perform a laboratory 
experiment where participants are asked to evacuate a place through a narrow 
door. They measured the evacuation time for 3 rushing levels of the 
participants and 6 different door widths (varying from 60\,cm to 
120\,cm). Afterward, they successfully calibrated the most sensitive parameters 
$\tau$ and $k_t$ to fit the experimental data. \\

In Ref.~\cite{lee2020speed}, Lee et al. perform an evacuation 
laboratory experiment varying the proportion of rush/no-rush participants. They 
used the Wasserstein distance metric to obtain the parameters $A$ and $B$ that 
best fitted the experimental results. They found many different sets of values 
$A$ and $B$ depending on the proportion of rush/no-rush participants. For the 
porpose of this work, we only selected the parameter values corresponding to 
100\% rush individuals.\\

The work of Frank et al. uses a subtle modification of the parameters of the 
model. They omit the body force ($k_n=0$) while fixing the other parameters with 
the same values proposed by Helbing et al.~\cite{helbing_2000}. The subject of 
study of this research is the emergency evacuations in the presence of 
obstacles~\cite{dorso_2011}. \\

In the research of Tang et al.~\cite{tang2011approach}, the authors perform 
pedestrian tracking in a Beijing metro station. They use a regression approach 
based on the least square method to fit the parameters $A$ and $B$. They obtain 
two sets of values depending on whether pedestrians are in an ``approach'' 
situation or a ``leave'' situation~\cite{tang2011approach}. The first one 
denotes pedestrians getting 
close to each other while the latter means pedestrians detaching from each 
other. We are only interested in the ``leave'' approach since we 
only consider repulsive interaction between pedestrians.\\

In the research of Sticco et al.~\cite{sticco2020effects}, The authors explore 
the role of the body force, the friction force, and their dynamical 
consequences in bottlenecks and corridors. They increased the parameter $k_t$ 
with respect to the original version of the SFM. The increment was done in order 
to reproduce the qualitative behavior of the fundamental diagram obtained from 
the 
measurements performed in Ref.~\cite{helbing2007dynamics}. In the research 
from Ref.~\cite{helbing2007dynamics}, the authors analyze a video recording of 
the massive muslim pilgrimage at the entrance of the Jamaraat bridge.\\

We list in Table~\ref{table_parameters} the sets of parameters that were 
selected for this research. Notice that these correspond to the set of 
parameters proposed/used in the works mentioned above. The list of parameters 
explicitly excludes the mass, radius, and desired velocity because we will set 
those parameters according to a different criterion from the ones proposed by 
the authors. In this work, we set the mass value $m=79.5\,$kg. For the 
pedestrian radius, we assigned two gaussian distributions: 
 $\mathcal{N}$(37.7\,cm,0.09\,cm) for females and 
$\mathcal{N}$(41.8\,cm,0.1\,cm)  for males. See section~\ref{numerical} for the 
details.\\

\begin{table}
\hspace*{-1cm}\begin{tabular}{c@{\hspace{6mm}}c@{\hspace{6mm}}c@{\hspace{6mm}}c@
{\hspace{6mm}}
c@{\hspace{6mm}}
c@{\hspace{14mm}}l}
 \hline
 Authors & $A\,$(N)   & $B\,$(m)     & $k_n\,$(Kg$\,$s$^{-2}$)  &  
$k_t\,$(Kg$\,$m$^{-1}\,$s$^{-1}$)  & $\tau\,$(s) &  Refs.\\
 \hline
Helbing et al.   & 2000    & 0.08 & $1.2\times 10^{5}$ & $2.4\times 10^{5}$ & 
0.50 & \cite{helbing_2000} \\
Li et al.  & 998    & 0.08 & 819 & 510 & 0.50 & \cite{li_2015} \\
Haghani et al. & 2000*    & 0.08* &  $1.2\times 10^{5}$* & 5500 & 0.12 & 
\cite{haghani2019simulating} \\
Lee et al.   & 2600    & 0.012 & 750 & 3000 & 0.50 & \cite{lee2020speed} \\
Frank et al.  & 2000    & 0.08 & 0 & $2.4\times 10^{5}$ & 0.50 & 
\cite{dorso_2011} \\
Tang et al.   & 729**     & 0.10 & $1.2\times 10^{5}$* & 
$2.4\times 
10^{5}$* & 0.60 & \cite{tang2011approach} \\
Sticco et al.   & 2000    & 0.08 & $1.2\times 10^{5}$ & $1.2\times 
10^{6}$ & 0.50 & \cite{sticco2020effects} \\

\hline \end{tabular} \caption{Sets of parameters selected for this research. 
Each row corresponds to a different set proposed/used in the literature. The 
cells with an asterisk * correspond to values that were completed with the 
Helbing's proposed values. The cell with ** is actually 9.18~$\times\,m$ with 
$m=79.5$.}
\label{table_parameters}
\end{table}

There are many similar works that we did not select because they calibrate 
other versions of the model (different from the escape panic 
version)~\cite{kabalan2016crowd,hoogendoorn2007microscopic,
seer2014validating,johansson2007specification,steiner2007parameter,
siddharth2018modeling, zeng2017specification,luber_2010,corbetta2015parameter}. 
The one in Ref.~\cite{kretz2015oscillations} studies the interval of 
parameters $A$ and $B$ that avoid oscillations in the model. This paper was 
excluded because the author explicitly avoids the bumping (and overlapping) in 
his theoretical derivation. This could be interpreted as an improvement of the 
model in the low-density limit (which is not the case for high-densities 
reported in emergency evacuations). The novel work of 
Bode~\cite{bode2020parameter} calibrates the escape panic version of the SFM 
using the experimental data from Ref.~\cite{experimental}. Nevertheless, this 
paper does not aim to provide a set of parameters but to promote a robust method 
for model calibration.\\

\section{\label{scope} Scope of this research and clarifications}

This investigation has two main goals. The first goal is to present the results 
of a novel empirical measurement from a real-life video analysis. The second 
goal is to compare the empirical measurements with numerical simulations. We 
will mention the strengths and weaknesses of the sets of parameters proposed by 
different authors across the literature. We do not pretend to provide an 
exhaustive description of all the phenomena presented here, but to explain the 
more salient features. We stress that this investigation is a call to set the 
SFM parameters with caution, rather than offering updated values for these 
parameters. We realize that more empirical data is required before we can assess 
definite estimates on this matter. \\

We based the research on the hypotheses that the video analyzed resembles an 
emergency evacuation. For simplicity reasons, we will refer to the rush to enter 
the Nike store as an evacuation process (although in evacuations, pedestrians 
feel the urge to escape from a particular place rather than being urged to enter 
into a building).\\

To simplify the analysis, we set the same parameter values to all the simulated 
agents (except for the pedestrian's diameter). We are aware that the reality is 
more complex, but we decided to leave this kind of analysis to future 
research.\\

In section \ref{optimization}, we optimize two parameters of the model ($k_n$ 
and $k_t$). We decided to focus on these two parameters because they are 
critical in high-density scenarios. The optimization of parameters like $A$ and 
$B$ (affecting the social repulsion) can be achieved in low-density situations. 
Whereas the optimization of $k_n$ 
and $k_t$ requires physical contact among pedestrians (like in the analyzed 
video). No further questioning will be done on the mutual relation between 
parameters in the reduced-in-units context of the SFM. This is out of the scope 
of the investigation. \\

\section{\label{empirical_numerical}Empirical measurements and numerical 
simulations}

In this section we describe the process undertaken to accomplish the empirical 
measurements. We also explain the details of the numerical simulations that we 
performed.\\

\subsection{\label{empirical}Empirical measurements}

The core of this work is the analysis obtained from a human rush event video. 
The video is available on YouTube~\cite{video} and shows a crowd of pedestrians 
rushing to 
enter a Nike store during a Black Friday event. The event took place at 
East Point Shopping Centre in the Ekurhuleni municipality (South Africa).\\

\begin{figure}[!htbp]
\centering
\subfloat[]{\includegraphics[trim=0 -1.9cm 0 0, width=0.49\columnwidth] 
{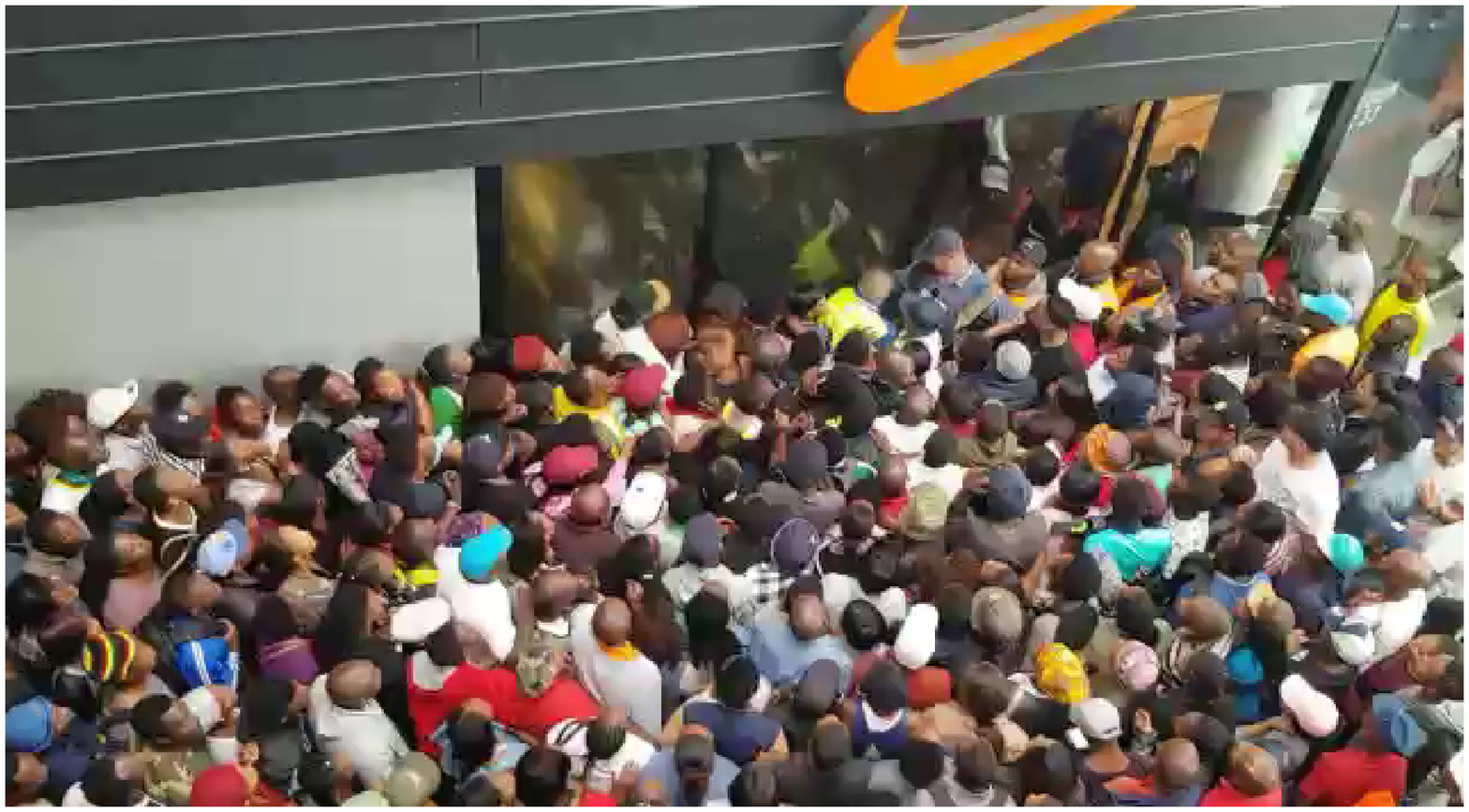}\label{snapshot_empirical}}\ 
\subfloat[]{\includegraphics[width=0.49\columnwidth]
{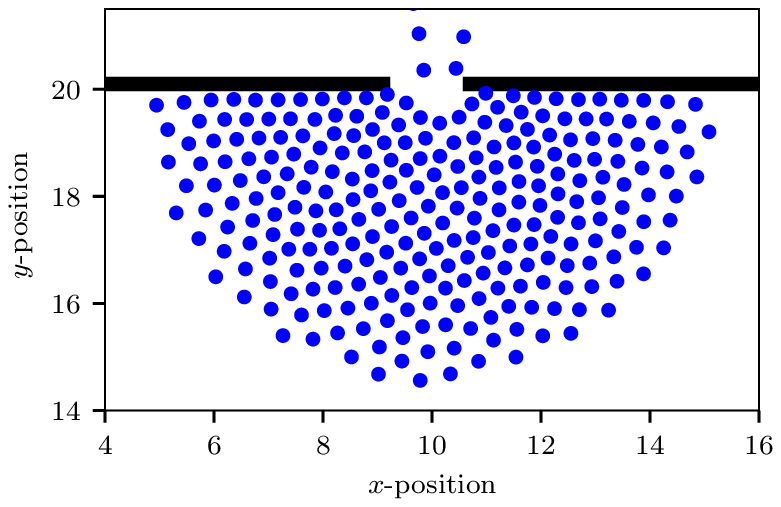}\label{snapshot_simulation}}\\
\caption[width=0.49\columnwidth]{\textbf{(a)} First frame of the analyzed 
video. A big crowd of pedestrians is close to the entrance of the store. The 
crowd is waiting for the security personnel to open the door. \textbf{(b)} 
Snapshot of the performed numerical simulations. The numerical simulations 
recreate the real-life situation from the video.}
\label{black_friday}
\end{figure}

Fig.~\ref{snapshot_empirical} is a snapshot of the first frame of the video. It 
shows a crowd of pedestrians waiting for the security personnel to allow them 
to 
get in. In the subsequent frames (see the video~\cite{video}), it is possible 
to see one 
person wearing a red t-shirt and white cup that manages to sneak into the store. 
Immediately after this, the crowd starts pushing towards the door, the security 
personnel steps aside, and the crowd begins to enter massively. Only at the 
end of the video, the pedestrians start to enter in a non-rushing fashion.\\

The door's capacity is exceeded by the large number of pedestrians trying to 
enter 
the store at the same time. The glass panel placed at the left side of the door 
is cracked as a result of the high pressures it supports. The pedestrians 
exhibit high competitive, aggressive, and non-cooperative behavior. We 
hypothesize that this situation could be similar to an emergency evacuation 
where the crowd is under panic, and self-interest may prevail over rational and 
collaborative behavior.\\

The video is composed by 1848 frames. We excluded the first 219 frames and the 
last 458. The first part was excluded because the crowd is waiting to 
enter the store. In a similar way, the last part was excluded because the crowd 
starts to enter in a much orderly fashion, and no competitive behavior is 
observed. The analysis 
was done from the frame 220 to the frame 1419 (including the border values). 
This is 1200 frames, equivalent to 40\,s of recording time (since the video 
frequency is 30 frames/s).\\

The selected 1200 frames correspond to the part of the video where the most 
competitive behavior was observed. Recall that we are only interested in 
scenarios like this, because the escape panic version of the SFM aims to 
reproduce emergency evacuations. The selected 1200 frames were divided into 
``segments'' of 60 frames each (equivalent to a sampling period of 
$\tau_{samp}=2\,$s). The criteria for choosing this sampling period was to be 
large enough to 
count a significant number of pedestrians who enter the store but, to be small 
enough to achieve a reasonable amount of measurements.\\

Once we divided the selected 1200 frames into 20 segments of 60 frames each, we 
proceed to count the number of ingressed pedestrians in each of the 20 
segments. \\

The video's quality is good enough to distinguish the pedestrians involved in 
the event. Therefore we did the counting manually frame-by-frame. No pedestrian 
tracking software was employed to achieve these measurements. In each frame, we 
draw a straight line that divided the inside of the store from the outside. As 
the video recording is almost steady, we only moved that dividing line once.  
Then, we counted how many pedestrians crossed the line in each segment of the 
video. \\

Once we counted how many pedestrians entered the store in each segment, we were 
able to calculate the mean flow (averaging) and the cumulative number of 
evacuees vs. time (evacuation curve).\\

\subsection{\label{numerical}Numerical simulations}

Since the escape panic version of the SFM aims to simulate emergency 
evacuations, we decided to recreate the conditions of the empirical measurements 
to test the SFM and different sets of parameters proposed in the literature.\\

The SFM was implemented on the LAMMPS simulation software \cite{plimpton}. 
Additional modules for LAMMPS were written in C++ to expand the software 
capabilities. To recreate the Black Friday event conditions, we set a door size 
of width $w=1.6\,$m. We set this value since it is almost equivalent to twice 
the size of a South African average door~\cite{door} (notice that the entrance 
of the store is a two-leaf door). Another criteria that made us choose 
$w=1.6\,$m is that the door is almost equivalent to 4 pedestrians' width 
(approximately 1.6\,m).\\

The total number of pedestrians was fixed as $N = 303$. The number $N$ is 
defined as $N=n+m$. Where $n=268$ is the total number of pedestrians that 
entered the store, while $m=35$ is the number of pedestrians that 
appear close to the door in the last analyzed frame. These $m$ 
pedestrians get inside the store a few frames after the last analyzed frame.\\

We set the simulated agents' diameter following the 
Ref.~\cite{mcdowell2009anthropometric}. We used the shoulder width 
(biacromial breadth) of adults corresponding to the ethnic group of 
``non-Hispanic black''. In the video, it is possible to observe that there is 
roughly an equal proportion of males and females. Therefore, we set half of the 
simulated agents as females and the other half as males. For each half, we set a 
gaussian distribution of shoulder widths. For males we assigned 
$\mathcal{N}$(41.8\,cm,0.1\,cm) and for females we assigned 
$\mathcal{N}$(37.7\,cm,0.09\,cm)~\cite{mcdowell2009anthropometric}.\\

We set a mass of $m=79.5\,$kg to all the simulated agents. This value is the 
average weight of men and women reported in  
Ref.~\cite{mcdowell2009anthropometric}. We tested whether variations in the mass 
value and diameter value could change the results. We found that the model is 
robust to mild variations in mass and diameter. \\

The initial conditions are configurations of agents that are located around the 
door (to reproduce the first frame of the video). To create these initial 
configurations, the $N$ agents were placed in a square enclosure with 
random positions and speeds. Then they were made to go to the door area but with 
the door closed (so that they do not leave the enclosure). After 20\,s the 
individuals form a semicircle around the door. That configuration is saved as an 
initial condition. We repeated this process 50 times, varying the initial 
positions and speeds of the agents.\\

In each simulation process, we recorded the number of agents that entered the 
store every 0.05\,s. The simulation process finished when $n=268$ simulated 
agents entered the store (which is the total number of ingressed pedestrians 
in the empirical measurements). We explored the interval of desired velocities 
$2\,$m/s$\,\leq v_d \leq 5$\,m/s. This interval corresponds to anxious 
individuals (but not extremely fast runners). \\

We use different sets of parameters to perform numerical simulations. The sets 
of parameters that we used are shown in the Table~\ref{table_parameters}. For a 
given set of parameters and a given desired velocity, we run 50 iterations 
varying the initial conditions. The Eq.~(\ref{newton_ec})  was numerically 
integrated employing the velocity Verlet algorithm, with a timestep value 
$\Delta t=$10$^{-4}\,$s. \\

\section{\label{results}Results}

We present in this section the main results of our investigation. The section is 
divided into two parts. In the first part (section \ref{Testing}), we compare 
the empirical measurements with numerical simulations by using different 
sets of parameter available in the literature. In the second part (section 
\ref{optimization}), we present the results of the parameter optimization 
by means of a genetic algorithm.\\

\subsection{\label{Testing}Testing}

In this section, we present and discuss our empirical results, and we further 
compare the numerical simulations with these empirical measurements.\\

In the previous section, we described the video analysis procedure to obtain 
the empirical measurements. Using this data, we were able to plot the empirical 
evacuation curve, which is the number of people entering the store as a 
function of time (cumulative number of evacuees vs. time). \\

\begin{figure}[!htbp]
\centering
\subfloat[]{\includegraphics[width=0.49\columnwidth]
{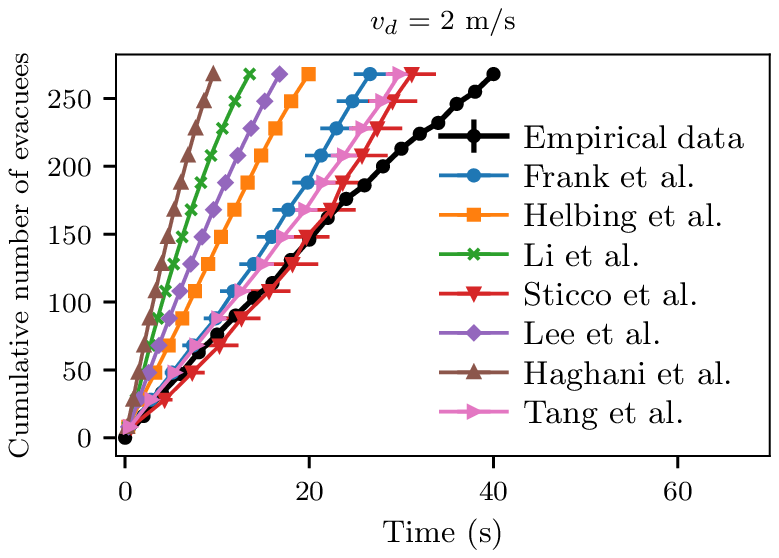}\label{discharge_vd2}}\ 
\subfloat[]{\includegraphics[width=0.49\columnwidth]
{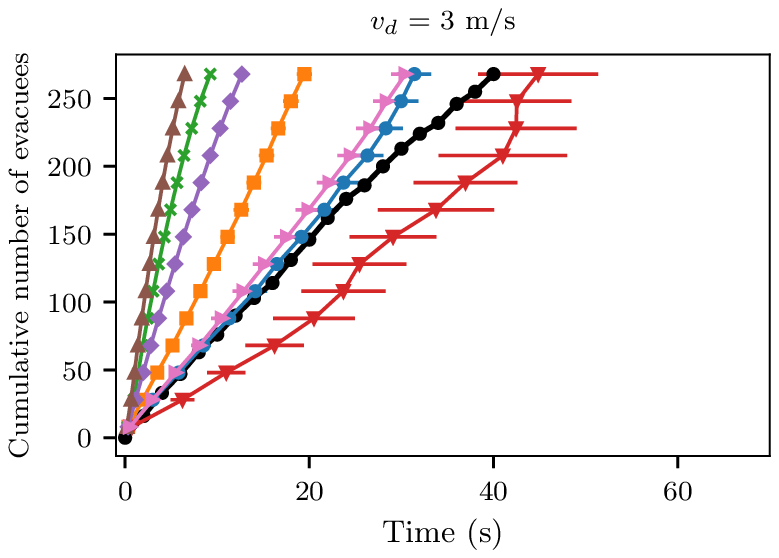}\label{discharge_vd3}}\\
\subfloat[]{\includegraphics[width=0.49\columnwidth]
{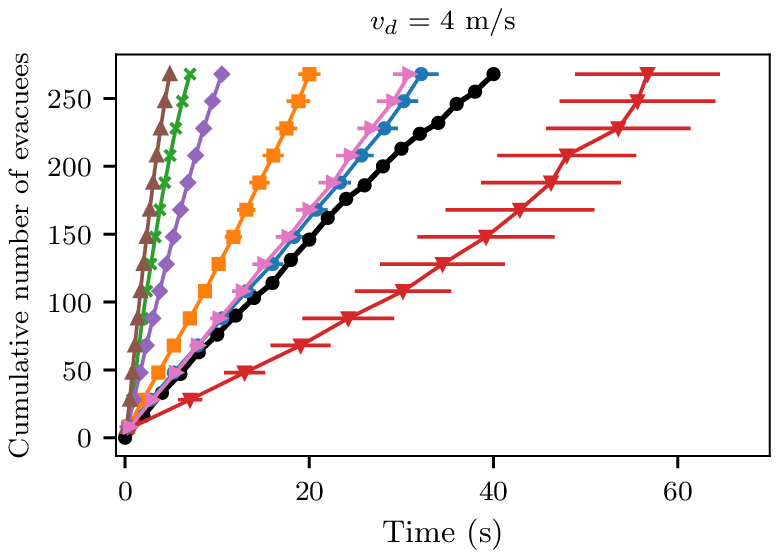}\label{discharge_vd4}}\ 
\subfloat[]{\includegraphics[width=0.49\columnwidth]
{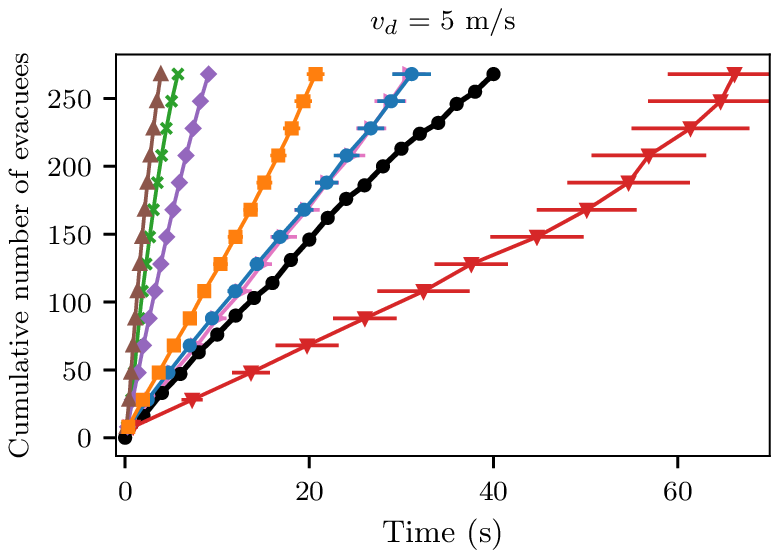}\label{discharge_vd5}}\\

\caption[width=0.47\columnwidth]{(color on-line only) Cumulative number of 
evacuees vs. time (evacuation curve). The black dotted curve corresponds to the 
empirical measurements; the rest of the curves correspond to numerical 
simulation results associated to the parameters proposed/used by different 
authors (see the legend and Table~\ref{table_parameters}). The simulation 
consisted of $N=$303 agents and finished when $n=$268 evacuated the place (to 
reproduce the empirical situation). The door's width was $w=1.6\,$m. The mass 
($m=79.5\,$kg) and shoulders width was set according to 
Ref.~\cite{mcdowell2009anthropometric}. 
50 iterations were run for each simulated curve, varying the initial 
conditions (the position and velocity of each agent). Each of the four figures 
corresponds to different desired 
velocities: (a),(b),(c),(d) correspond to $v_d=2\,$m/s, $v_d=3\,$m/s, 
$v_d=4\,$m/s and $v_d=5\,$m/s, respectively.}
\label{sim_vs_empirical}
 \end{figure}

The empirical evacuation curve is a monotonically increasing function (see the 
black dotted line in Fig.~\ref{sim_vs_empirical}). There are no time intervals 
in which the empirical evacuation process is completely stopped, at least for 
the sampling frequency employed in this research ($\tau_{samp}=2\,$s). We may 
speculate that a lower $\tau_{samp}$ would capture the stop-and-go mechanism 
produced by the blocking clusters close to the 
door~\cite{parisi2007morphological}.\\

We used the empirical measurements to calculate the mean flow 
of pedestrians entering the store  $\langle J  \rangle =6.7 \pm 0.8\,$p/s. The 
flow value is higher than other experimental measurements at bottlenecks where 
individuals are not allowed to harm one 
another~\cite{liddle2009experimental,kretz2006experimental,
muller1981gestaltung,muir1996effects,nagai2006evacuation,seyfried2007capacity,
nicolas2017pedestrian,garcimartin2016flow,jianyu2019experimental}.\\

The specific flow is defined as the flow scaled by the door 
size (say,  $\langle J  \rangle / w$). This quantity is used to compare 
evacuation processes with different 
door dimensions (to a certain extent). The mean specific flow corresponding to 
our measurement is $\langle J_s  \rangle =4.1 \pm 0.5\,$p/m.s  (assuming a door 
width of $1.6\,$m). To our knowledge, only the specific flow reported in 
Refs.~\cite{garcimartin2016flow,nagai2006evacuation} is comparable to ours. 
However, the specific flow from 
Refs.~\cite{garcimartin2016flow,nagai2006evacuation} corresponds to door sizes 
much narrower ($w\leq0.75\,$m) than the estimated door size of the Nike store 
($w=1.6\,$m). It is worth mentioning that some researches have reported that 
narrower doors produce a moderate increment in the specific 
flow~\cite{kretz2006experimental} while other researches concluded the opposite 
result~\cite{seyfried2009new}.\\

Fig.~\ref{sim_vs_empirical} shows the cumulative number of pedestrians that 
entered the store as a function of the time (for both empirical measurements 
and numerical simulations). Fig.~\ref{sim_vs_empirical} is a collection of four 
figures, all of them include the empirical result (shown as a black dotted 
line) and the results of the numerical simulations (shown as colored curves). 
Each of these curves is associated with different sets of parameters 
(corresponding to the selected papers mentioned in 
Table~\ref{table_parameters}).\\

Each figure from Fig.~\ref{sim_vs_empirical} is 
associated to a different desired velocity in the interval 
$2\,$m/s$\,\leq v_d \leq 5$\,m/s. The numerical results of 
Fig.~\ref{discharge_vd2} and 
Fig.~\ref{discharge_vd3} correspond to $v_d=2\,$m/s and $v_d=3\,$m/s,  while 
the numerical results of Fig.~\ref{discharge_vd4} and Fig.~\ref{discharge_vd5} 
correspond to $v_d=4\,$m/s and $v_d=5\,$m/s, respectively. Recall that we assign 
the same desired velocity to every simulated agent from the beginning of the 
simulation until the simulated agent manages to enter the store.\\ 

The evacuation time $t_e$ is defined as the time for which all the pedestrians 
evacuated the place (for example, the evacuation time for the empirical 
measurements is $t_e = 40\,$s). The Faster-is-Slower is a phenomenon that 
occurs when increasing the desired velocity $v_d$ reduces the evacuation time. 
The Faster-is-Faster is the opposite effect; this means that the higher the 
value of $v_d$, the lower the evacuation time. Within this framework, we can 
distinguish two categories of evacuation curves: the curves that exhibit 
Faster-is-Faster (Lee, Li, Haghani) and the curves that exhibit Faster-is-Slower 
(Sticco, Frank, Tang). The curves corresponding to the parameters proposed by 
Helbing et al. do not display any of these effects significantly.\\

Within the three sets of parameters that exhibit Faster-is-Slower, the most 
noticeable curves are the corresponding to the parameter set proposed by Sticco 
et al. Recall that this parameter set is similar to Helbing's but with the 
friction coefficient increased by a factor of five. This correction was 
suggested in order to reproduce the qualitative behavior of the fundamental 
diagram at the entrance of the Jamaraat bridge~\cite{sticco2020effects}. The 
evacuation curve shown in 
Fig.~\ref{discharge_vd3} (for $v_d=2\,$m/s), seems to attain a good agreement 
with the empirical 
measurements until time $t \simeq 20$\,s. Above this time, the simulated curve 
tends to increase faster than the empirical curve. For higher desired 
velocities ($v_d \geq 3\,$m/s), the parameters proposed by Sticco et al. 
produce 
very large evacuation times due to the high friction value. \\

The parameter sets proposed by Frank et al. and Tang et al. produce a mild 
Faster-is-Slower effect. They seem to be the parameter sets that produce the 
best agreement with this empirical data. Notice that the evacuation times 
($i.e$ the time corresponding to the last measurement) surpass 
the evacuation time from Helbing et al. (for any of the desired velocity 
explored). In the case of Frank et al., this difference can be explained because 
the body force is not considered ($k_n=$0), this produces a more significant 
overlap between pedestrians, which ultimately leads to an increment in 
the friction~\cite{sticco2020effects}.\\

In the case of the parameters proposed by Tang et al., it is more difficult to 
compare them with Helbing's because three parameters are modified ($A$, $B$, 
and $\tau$). Nevertheless, the value of $A$ is reduced and the value of $B$ is 
increased. These changes tend to diminish the social force repulsion, which 
presumingly leads to an effective increment of the overlap (and the friction 
force). This is analogous to the $k_n$ reduction stated by Frank et al.\\

As we mentioned before, the parameters proposed by Haghani, Lee and Li yield 
Faster-is-Faster effect. The three of them produce evacuation times below the 
empirical measurements (and also below the evacuation time corresponding to 
Helbing). This result holds true for any of the explored desired velocities.\\

The set of parameters proposed by Haghani et al. and Lee et al. were able to 
reproduce the experimental results in laboratory conditions where pedestrians 
are not allowed to have aggressive behavior against each other. The Black 
Firday event (empirical condition) reported in this paper involves highly 
aggressive and competitive behavior (see the video from Ref.~\cite{video}). 
This discrepancy between laboratory and empirical conditions seems to be the 
reason why the sets of parameters proposed in 
Refs.~\cite{haghani2019simulating,lee2020speed} do not 
produce evacuation curves in agreement with the empirical data analyzed in this 
paper.\\

The set of parameters proposed by Li et al. was calibrated using a real-life 
(empirical) video analysis. The situation analyzed is an emergency evacuation of 
a classroom due to the 2013 Ya\textquotesingle an earthquake in China. In the 
video~\cite{video_yaan}, it is possible to observe that students rush to 
evacuate the classroom. Despite this, the physical contact between the students 
involved is almost negligible. This particular trait of the Ya\textquotesingle 
an earthquake evacuation may have underestimated the value of the friction 
coefficient $k_t$, and consequently, the expected impact that $k_t$ could have 
in a high-density situation such as the Black Friday event analyzed in this 
work.\\

We may summarize this Section as follows. We reported the evacuation curve of 
the Black Friday event and measured the flow $\langle J  \rangle =6.7 \pm 0.8 
\,$p/s (and estimated the specific flow $\langle J_s  \rangle =4.1 \pm 0.5 
\,$p/m.s). This is a higher value than the flow reported under laboratory 
conditions throughout the literature. We performed numerical simulations to 
reproduce the Black Friday event's conditions using different sets of parameters 
available in the literature. We divided the results into two categories 
(parameter sets that yield Faster-is-Slower pattern and parameter sets that 
produce Faster-is-Faster pattern). The 
parameter sets that yield Faster-is-Slower seem to have better agreement with 
the empirical curve (except for the parameter set of Sticco et al. in the 
interval $v_d \geq 3\,$m/s). The sets of parameters that yield 
Faster-is-Faster 
were calibrated from laboratory and empirical situations. In these situations, 
the pedestrians involved do exhibit neither aggressive behavior nor significant 
physical contact. This fact may have underestimated the value of the friction 
coefficient $k_t$, leading to evacuation curves with evacuation times lower than 
the empirical case. \\

\subsection{\label{optimization}Parameter optimization}

In this section, we present the results of the parameter optimization. We 
perform an optimization of the parameters $k_n$ and $k_t$; these are the 
parameters associated with the physical forces: the body force and the 
friction. The 
optimization was done using the Differential Evolution (DE) algorithm, which is 
a technique that belongs to the family of genetic algorithms (see 
Ref.~\cite{storn1996usage} for a detailed explanation of the algorithm). \\

The goal of DE is to optimize the parameters of the model in order to achieve an 
evacuation curve consistent with the empirical measurements obtained from the 
Black Friday video. The optimization process finished when the algorithm reached 
100 generations. Since we aim to optimize two parameters, we set 2$\times 10 =$ 
20 different sets of random $(k_n,k_t)$ values following the recommendations of 
Storn~\cite{storn1996usage}. The crossover probability and the differential 
weight were set as $CR=0.3$ and $F=0.5$, respectively.\\

The algorithm requires an objective function (a function to minimize), we 
defined this function as 

\begin{equation}
f(k_n,k_t)=\frac{\sum_{i}^{M}  \left | t_i^s(k_n,k_t)-t_i^e \right |}{M}
\label{objective_function}
\end{equation}

\noindent where $t_i^s(k_n,k_t)$ is a time value corresponding to the simulated 
evacuation curve,  $t_i^e$ is a time value corresponding to the empirical 
evacuation curve and $M=21$ is the total number of measurements in the 
evacuation curve. The index $i$ is associated with a fixed number of evacuees. 
For instance, $i=0$ corresponds to zero evacuees, while $i=21$ corresponds to 
268 evacuees (both in the empirical curve and in the simulated curve). 
$f(k_n,k_t)$ is a real number that expresses how much the simulation differs 
from the empirical measurement. Notice that this estimator 
avoids the usual overweighting of large deviations encountered in quadratic 
estimators.  \\

The numerical simulations have the same characteristics as the simulations 
from the previous section (total number of evacuees, door size, shoulders width, 
etc.). The only two parameters that we aim to optimize are the friction 
coefficient $k_t$ and the body force parameter $k_n$. The other parameters were 
fixed to the values proposed by Helbing et al.~\cite{helbing_2000} (except for 
the pedestrian mass and radius).\\

\begin{figure}[!htbp]
\centering
\subfloat[]{\includegraphics[width=0.49\columnwidth]
{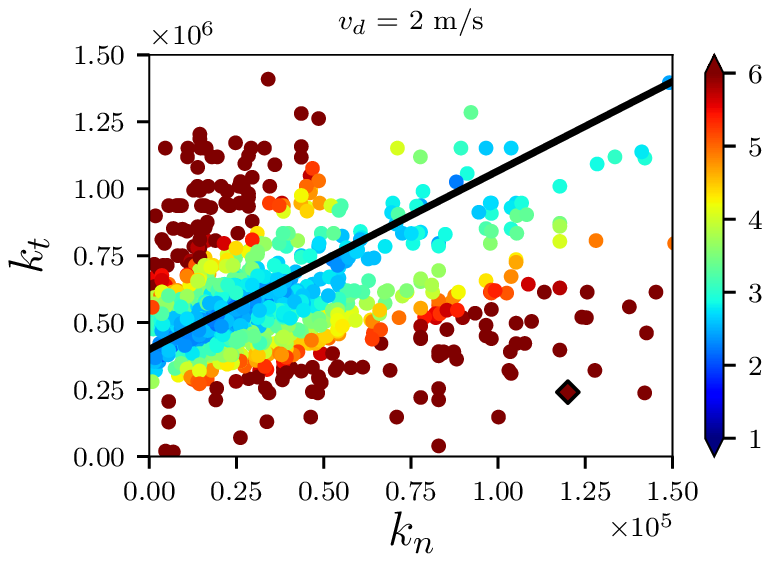}\label{landscape_vd2}}\ 
\subfloat[]{\includegraphics[width=0.49\columnwidth]
{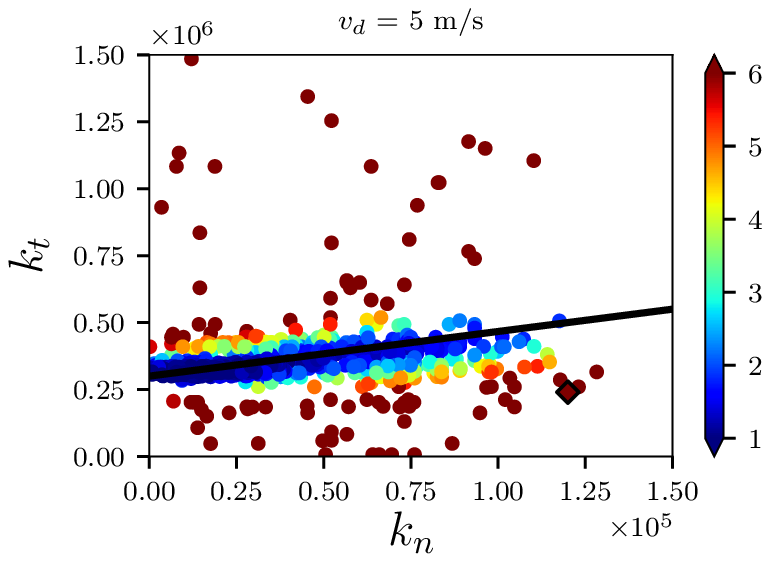}\label{landscape_vd5}}\\
\caption[width=0.49\columnwidth]{Fitness landscape for the parameters $k_n$ 
(horizontal axis) and $k_t$ (vertical axis). The rest of the parameters are the 
same as Helbing's proposal in Ref.~\cite{helbing_2000}. The mass ($m=79.5\,$kg) 
and shoulders width was set according to Ref.~\cite{mcdowell2009anthropometric}. 
The color scale represents the value of the objective function $f(k_n,k_t)$ (the 
result of comparing the numerical simulation against the empirical data). The 
circular dots were produced by the DE algorithm to minimize $f(k_n,k_t)$. The 
diamond dot represents the parameters of Helbing et al.~\cite{helbing_2000} 
\textbf{(a)} Corresponds to simulated agents with $v_d=2\,$m/s while 
\textbf{(b)} corresponds to simulated agents with $v_d=5\,$m/s.}
\label{landscapes}
\end{figure}

Given the values $(k_n,k_t)$, the algorithm calculates the mean evacuation curve 
over 50 iterations. The mean evacuation is then compared against the empirical 
evacuation curve (see Eq.~(\ref{objective_function})) to obtain a value 
$f(k_n,k_t)$. Fig.~\ref{landscapes} shows the 
fitness landscape, the horizontal axis stands for the parameter $k_n$ and the 
vertical axis stands for the parameter $k_t$. The color bar represents the value 
of the objective function $f(k_n,k_t)$.\\

Fig.~\ref{landscape_vd2} shows the fitness landscape for pedestrians with a 
desired velocity $v_d=2\,$m/s while Fig.~\ref{landscape_vd5} corresponds to 
pedestrians with $v_d=5\,$m/s. Both fitness landscapes show that the optimal 
values of the parameters $k_n$ and $k_t$ are correlated. The solid black line is 
a visual guide to see the correlation in the parameter space. In both plots, it 
is possible to observe that increasing the value of $k_n$ requires increasing 
the parameter $k_t$ to achieve a similar fitness value $f(k_n,k_t)$.\\

The correlation can be explained because the friction force depends on the value 
of $k_t$, but it also depends on the overlap between contacting pedestrians 
(see the first term in Eq.~(\ref{eqn_physical})). The 
value of $k_n$ affects the overlap value (which subsequently affects the 
friction). Many different combinations of the parameters may produce similar 
results (see Ref.~\cite{sticco2020re} for a further discussion). The black solid 
line has a different slope depending on the value of $v_d$. Increasing $v_d$ 
reduces the slope because $v_d$ directly affects the overlap between pedestrians 
(the higher the desired velocity, the higher the overlap). In other words, if 
the desired velocity is higher, the friction coefficient $k_t$ needs to be 
lower in order to achieve the same results.\\

\begin{figure}[!htbp]
\centering
\subfloat[]{\includegraphics[width=0.7\columnwidth]
{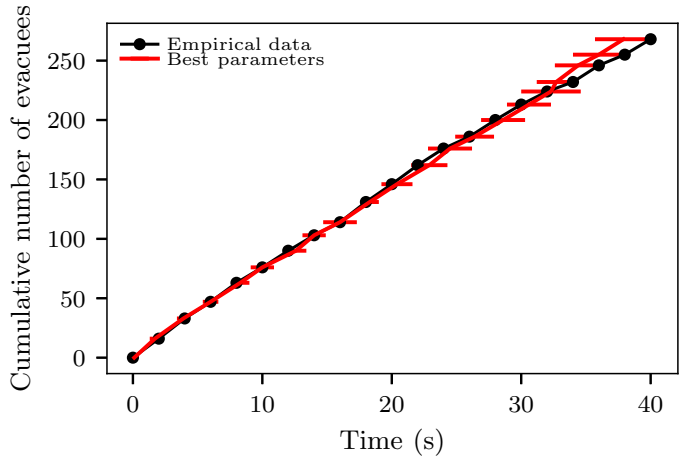}}\
\caption[width=0.49\columnwidth]{ Cumulative number of evacuees as a function 
of time (evacuation curve). The black dotted line corresponds to empirical 
measurements while the red line corresponds to the numerical simulation result 
that best reproduces the empirical data. The desired velocity is $v_d=5\,$m/s . 
The parameters of the best fit are $(k_n,k_t)=(0.0364\times10^{5},3.05\times10^ 
{5}$), the rest of the parameters are the same as Helbing's proposal in 
Ref.~\cite{helbing_2000}. The mass ($m=79.5\,$kg) and shoulders width was set 
according to Ref.~\cite{mcdowell2009anthropometric}.}
\label{fit}
\end{figure}

Fig.~\ref{fit} shows the empirical evacuation curve and the curve corresponding 
to the best set of parameters (red line) for $v_d=5\,$m/s. The best set of 
parameters was obtained with the DE algorithm. It corresponds to 
$f(k_n,k_t)=0.57$ and parameters: 
$(k_n,k_t)=$(0.036$\times10^{5}$, 3.05$\times10^{5}$). The friction coefficient 
is higher than the value corresponding to the Helbing's original 
proposal (which is $k_t=2.4\times10^{5}$). This result is consistent with the 
friction increment suggested in Ref.~\cite{sticco2020re}. On the other hand, the 
body force coefficient is lower than the original value 
($k_n=$1.2$\times10^{5}$). This value has already been questioned in 
Ref.~\cite{cornes2017high} since empirical data suggests a 
lower value for the human compression coefficient. However, we emphasize that 
there is no unique 
combination that produces the optimal result since many different parameter 
combinations can produce similar outcomes.\\

If the desired velocity was a known value, we could determine the combination of 
parameters ($k_n$ and $k_t$) that best fits the empirical data. Although we did 
not estimate the value of $v_d$, the fitness landscapes shown in this research 
can help to narrow down the possible parameters that could fit the empirical 
evacuation curve.\\

For future researches, it will be necessary to find a set of parameters that
are able to fit situations like the Black Friday event and also situations such 
as the Jamaraat pilgrimage (where $v_d$ is low, but the density is extremely 
high). In the latter case, it would be convenient to fit the parameters using 
the fundamental diagram measurements reported in 
Ref.~\cite{helbing2007dynamics}.

We conclude this second part of the results by mentioning that the Differential 
Evolution algorithm could achieve a set of parameters for $v_d=5\,$~m/s
$(k_n,k_t)=(0.0364\times10^{5},3.05\times10^{5})$ that fits the 
empirical data with an agreement value of $f(k_n,k_t)=0.57$. This result 
suggests that the friction value should be increased respect the original value 
proposed in Ref.~\cite{helbing_2000}. We also stress that there 
is no unique combination of optimal parameters but a collection of possible 
combinations that lead to similar results in terms of the objective function.\\

\section{\label{conclusions}Conclusions}

The escape panic version of the Social Force Model has been widely considered in 
pedestrian dynamics. The purpose of the model is to simulate emergency 
evacuations under high emotional stress. We have 
analyzed a real-life video (Black Friday event) that resembles an emergency 
evacuation and then turns out to be suitable for testing and optimizing the 
model.\\

We measured the Black Friday event flow ($\langle 
J  \rangle =6.7 \pm 0.8 \,$p/s) and found that it is higher than the flow 
measurements reported in most experimental 
studies. We argue that this 
discrepancy is related to the fact that in controlled experiments, the 
participants are not extremely competitive (due to safety considerations). On 
the other hand, in the real-life situation analyzed in this paper, it is 
possible to observe a high level of competitiveness and aggression, which may 
have lead to such a distinct result. \\ 

We performed numerical simulations that recreate the Black Friday event. We 
used different sets of parameters that were previously proposed/used in the 
literature. The parameter sets calibrated from laboratory controlled 
experiments 
or situations where the physical contact is negligible produced simulations 
which display a strong disagreement with our empirical measurements. These sets 
of parameters produce simulated agents that evacuate too fast. We think that 
even though those 
calibrations fit well the situations analyzed in their researches, they are not 
completely suitable for analyzing high stress situations like the Black Friday 
event because the friction contribution turns out to be underestimated. \\

The sets of parameters from Refs.~\cite{dorso_2011,tang2011approach} provide 
better reproduction of the empirical results, although they do not fully 
reproduce the empirical evacuation curve of the Black Friday event.\\

In order to explore the possibility of getting better values of the parameters 
$k_n$ and $k_t$, we implemented a Differential Evolution 
algorithm to optimize them. The optimization criteria was to minimize the 
difference between the simulated evacuation curve and the empirical evacuation 
curve. The values that produce the best fit 
are: $(k_n,k_t)=(0.0364\times10^{5},3.05\times10^{5}$), for $v_d=5\,$m/s. These 
values reproduce the qualitative and quantitative behavior of the empirical 
measurement. We also found that $k_n$ and $k_t$ are correlated in the fitness 
landscape. The consequence of this is that there are multiple combinations of 
$k_n$ and $k_t$ that produce the same fitness values. We did not attempt to 
identify the source of correlation, but we are currently working on this topic 
for an upcoming investigation.\\

We stress that physical forces (friction and body force) are a 
critical factor in reproducing emergency evacuations where the pedestrians are 
subject to high-density and high anxiety conditions. Therefore, we suggest 
calibrating the model parameters using real-life emergency evacuations or 
similar situations (like the Black Friday stampede analyzed in this paper). The 
results obtained after performing the above mentioned optimization, indicate 
that the friction coefficient $k_t$ is generally underestimated.\\

We believe that in the coming years, there will be an increasing number of 
videos like the one analyzed in this work. Situations like Black Friday events, 
music concerts, and human stampedes are more often documented and published on 
the internet. This data, together with computer vision techniques, will 
contribute to study these singular events. We hope that future researches based 
on real-life video analysis will improve the theoretical models aimed at 
simulating emergency evacuations. \\

\section*{Acknowledgments}

This work was supported by the National Scientific and Technical 
Research Council (spanish: Consejo Nacional de Investigaciones Cient\'\i ficas 
y T\'ecnicas - CONICET, Argentina) grant Programaci\'on Cient\'\i fica 2018 
(UBACYT) Number 20020170100628BA.\\

G.A Frank thanks Universidad Tecnol\'ogica Nacional (UTN) for partial
support through Grant PID Number SIUTNBA0006595.\\

\end{document}